\documentclass[iop,apj]{emulateapj}
\usepackage{natbib}
\usepackage{xspace}
\usepackage{xcolor}
\bibpunct{(}{)}{;}{a}{}{,} 
\usepackage[
  pdfstartview=FitH,
  linkcolor=blue,
  anchorcolor=blue,
  citecolor=blue,
  urlcolor=blue,
  colorlinks=true,
  breaklinks]{hyperref}

\newcommand{\alphat}{\ensuremath{\alpha_\mathrm{t}}\xspace}
\newcommand{\cs}{\ensuremath{c_\mathrm{s}}\xspace}
\newcommand{\cc}{\ensuremath{c_\mathrm{c}}\xspace}

\newcommand{\eps}{\ensuremath{\epsilon}\xspace}

\newcommand{\Ok}{\ensuremath{\Omega_\mathrm{k}}\xspace}

\newcommand{\rc}{\ensuremath{r_\mathrm{c}}\xspace}

\newcommand{\rhos}{\ensuremath{\rho_\mathrm{s}}\xspace}
\newcommand{\Sigd}{\ensuremath{\Sigma_\mathrm{d}}\xspace}
\newcommand{\Sigg}{\ensuremath{\Sigma_\mathrm{g}}\xspace}
\newcommand{\Signull}{\ensuremath{\Sigma_0}\xspace}
\newcommand{\St}{\ensuremath{\mathrm{St}}\xspace}
\newcommand{\Tc}{\ensuremath{T_\mathrm{c}}\xspace}
\newcommand{\taug}{\ensuremath{\tau_\mathrm{g}}\xspace}
\newcommand{\Vc}{\ensuremath{V_\mathrm{c}}\xspace}
\newcommand{\Vk}{\ensuremath{V_\mathrm{k}}\xspace}
\newcommand{\figscalefact}{1.13}
\makeatletter
\newif\ifhbonecolumn
\@ifclasswith{emulateapj}{onecolumn}{\hbonecolumntrue}{\hbonecolumnfalse}
\newif\ifbaastex
\@ifclasswith{aastex}{iop}{\baastextrue}{\baastexfalse}
\makeatother
\newcommand{\foo}{
    \ifhbonecolumn
        \fontsize{12}{20}\selectfont
        \renewcommand{\figscalefact}{0.6}
    \else
    \fi
}
\newcommand{\figscale}[1]{
    \ifbaastex
        \epsscale{#1}
    \else
        \epsscale{\figscalefact}
    \fi
}

\shorttitle{Outer Edges of Dust Disks}
\shortauthors{Birnstiel \& Andrews}
\begin{document}
\title{On the Outer Edges of Protoplanetary Dust Disks}
\author{Tilman Birnstiel and Sean M. Andrews}
\affil{Harvard-Smithsonian Center for Astrophysics, 60 Garden Street, Cambridge, MA 02138}
\email{tbirnstiel@cfa.harvard.edu, sandrews@cfa.harvard.edu}

\begin{abstract}
The expectation that aerodynamic drag will force the solids in a gas-rich protoplanetary disk to
spiral in toward the host star on short timescales is one of the fundamental problems in planet
formation theory. The nominal efficiency of this radial drift process is in conflict with
observations, suggesting that an empirical calibration of solid transport mechanisms in a disk is
highly desirable. However, the fact that both radial drift and grain growth produce a similar
particle size segregation in a disk (such that larger particles are preferentially concentrated
closer to the star) makes it difficult to disentangle a clear signature of drift alone.
We highlight a new approach, by showing that radial drift leaves a distinctive ``fingerprint'' in
the dust surface density profile that is directly accessible to current observational facilities.
Using an analytical framework for dust evolution, we demonstrate that the combined effects of drift
and (viscous) gas drag naturally produce a sharp outer edge in the dust distribution (or,
equivalently, a sharp decrease in the dust-to-gas mass ratio). This edge feature forms during the
earliest phase in the evolution of disk solids, before grain growth in the outer disk has made much
progress, and is preserved over longer timescales when both growth and transport effects are more
substantial. The key features of these analytical models are reproduced in detailed numerical
simulations, and are qualitatively consistent with recent millimeter-wave observations that find
gas/dust size discrepancies and steep declines in dust continuum emission in the outer regions of
protoplanetary disks.
\end{abstract}


\foo


\section{Introduction}\label{sec:intro}

Much of the current research on circumstellar disks is focused on forging links to planet formation,
specifically through difficult observational examinations of the disk material within a few tens of
AU from its host star. By comparison, the outer reaches of these disks ($\sim$hundreds of AU)
receive little attention, despite being more easily accessible to current telescopes (especially at
mm/radio wavelengths). Observations of the material at large disk radii are indirectly quite 
relevant to the planet formation process: the outer disk contains the majority of the mass reservoir
available for making planets, and resolved measurements of its structure can potentially reveal
clues to some fundamental disk evolution mechanisms
\citep[e.g.,][]{Andrews:2009p7729,Andrews:2010p17519}.

In the context of the outer disk, a \textit{size} seems like a basic property that can be easily
inferred. However, resolved observations indicate that such a measurement is problematic: the
apparent size depends on the adopted tracer. Radio interferometer data suggest that line emission
from abundant molecules (e.g., CO) appears more spatially extended than continuum radiation from
dust. In the initial studies of outer disk structures, this discrepancy was dismissed as an artifact
of limited continuum sensitivity \citep{Dutrey:1998p21054,Guilloteau:1998p21055}. The emission line
tracers of molecular gas are highly optically thick, so even a small amount of gas far from the host
star still emits at a detectable level \citep[e.g.,][]{Beckwith:1993p21052}. But, the continuum
emission at these wavelengths is optically thin \citep{Beckwith:1990p3768}, so the corresponding
radiation from dust was much too faint for those early observations. However, that explanation
proved fleeting when the observed gas/dust size discrepancy persisted even after the sensitivity
improved \citep{Pietu:2005p16795,Pietu:2007p4501,Isella:2007p21062}. An elegant solution proposed by
\citet{Hughes:2008p21059} showed that a more appropriate model for the density profile in a viscous
disk -- with an exponential taper at large radii -- reconciles the discrepancy (again through an
optical depth/sensitivity effect).

Yet, remarkably, the continued improvement in data quality has revealed that even this solution is
insufficient. The line/continuum size discrepancies remain an issue, leading some to speculate that
there is an intrinsic, physical difference in the radial distributions of gas and dust such that the
dust-to-gas mass ratio decreases with distance from the host star
\citep{Panic:2009p11789,Andrews:2012p16676,deGregorioMonsalvo:2013p21848,Rosenfeld:2013p21297}.
Moreover, in the best currently available datasets, the continuum emission is found to exhibit a
\textit{sharp} decrease over a narrow radial range \citep[$\Delta r/r \lesssim
0.1$;][]{Andrews:2012p16676,deGregorioMonsalvo:2013p21848}. This latter feature is tantalizingly
reminiscent of the precipitous drop in the population of large bodies in the classical Kuiper Belt
$\gtrsim$50\,AU from the Sun \citep{Jewitt:1998p21067,Allen:2001p21050,Trujillo:2001p21096}. Taken
together, these attempts to measure disk sizes have forced observers to suggest that the radial
profile of the dust-to-gas ratio features a steep decrease -- an ``edge" -- in the outer regions of
a protoplanetary disk. So far, a physical explanation for this feature is lacking, although it has
anecdotally been associated with a profound change in the grain properties.

In this article, we develop theoretical models which suggest that both the gas/dust size 
discrepancy and the sharp dust edges observed in disks at radii of tens or hundreds of AU are 
natural, generic consequences of the growth and migration of solids in a viscously evolving gas
disk. In Sections~\ref{sec:early_phase} and \ref{sec:late_phase}, we describe our calculations of
this evolution in some relevant analytical limits: an ``early" phase where an outer dust edge forms
rapidly before substantial particle size evolution, and a ``late" phase where particles have 
already reached a maximum equilibrium size. In Section~\ref{sec:comparison}, we verify these 
findings with more sophisticated numerical simulations that track the complete evolution, without 
these simplifying assumptions. Finally, in Section~\ref{sec:conclusion} we discuss the results and
highlight how the analysis of resolved continuum and molecular line observations of disks can help
empirically calibrate such particle evolution models in the near future.

\section{Early Phase}\label{sec:early_phase}

In this section, we consider the radial migration of dust grains in the outer regions of a
protoplanetary disk during an ``early" phase, where the evolution of particle sizes has not yet
progressed enough to significantly affect their dynamics. For simple assumptions about typical
collisions, the growth (or size-doubling) timescale can be expressed as
\citep{Kornet:2001p688,Brauer:2008p215}
\begin{equation}
\taug = \frac{1}{\eps\,\Ok\,S},
\label{eq:tau_grow}
\end{equation}
with $\eps = \Sigd/\Sigg$ the dust-to-gas surface density ratio, \Ok the Keplerian orbital
frequency, and $S$ the sticking efficiency (taken to be unity). In the outer disk and for the grain 
sizes of interest here, the main source of collision velocities is turbulence \citep[][but see 
Section \ref{sec:conclusion}]{Ormel:2007p801}. For typical outer disk parameters, estimates of
\taug (and the duration of this phase) are roughly 0.1\,Myr.

There are two primary radial transport mechanisms for dust embedded in a gas disk: (1)
\textit{radial drift}, caused by an angular momentum exchange between dust particles and the
sub-Keplerian orbital motion of the gas; and (2) \textit{gas drag}, produced by the coupling of dust
particles in the gas flows controlled by viscous evolution. The dust velocity imparted by drift is
\citep{Nakagawa:1986p2048}
\begin{equation}
u_\mathrm{rd} = \frac{1}{\St + \St^{-1}}\,\frac{\cs^2}{\Vk} \frac{\mathrm{dln}P}{\mathrm{dln}r},
\label{eq:radial_drift}
\end{equation}
with sound speed \cs, Keplerian velocity \Vk, and gas pressure $P$. And the dust velocity 
introduced by gas drag is
\begin{equation}
u_\mathrm{gd} = \frac{1}{1+\St^2}\,u_\mathrm{g},
\label{eq:u_gas_drag}
\end{equation}
with a gas radial velocity 
\begin{equation}
u_\mathrm{g} = -\frac{3}{\Sigg\,\sqrt{r}}\frac{\partial}{\partial
r}\left(\Sigg\,\nu\,\sqrt{r}\right),
\label{eq:u_gas}
\end{equation}
where we assume
\begin{equation}
\nu = \alphat \, \frac{\cs^2}{\Ok}
\end{equation}
as a parametrization of the gas viscosity \citep{Shakura:1973p4854}. In Eq.~\ref{eq:radial_drift},
we introduced the Stokes number \St, a dimensionless size that characterizes the coupling strength
between the dust and gas. The drag force is always in the Epstein regime for the regions of the disk
and the particle sizes of interest here. In that case, the Stokes number at the disk mid-plane can
be written as
\begin{equation}
\St \simeq \frac{a\,\rhos}{\Sigg}\,\frac{\pi}{2},
\label{eq:stokes_number}
\end{equation}
where $a$ is the grain radius and \rhos the internal density of the grain, assumed here to be 1.6 g
cm$^{-3}$.

To illustrate the dynamical behavior that these transport mechanisms have on dust particles, we need
to first adopt a set of representative parameters that characterize the disk structure. The gas
temperatures (i.e., sound speeds) influence the dust velocities, and the gas surface densities
affect the dust-gas coupling. Here and throughout this article, we assume that the gas surface
density profile can be described by the standard, self-similar solution to the viscous evolution
equations presented by \citet{LyndenBell:1974p1945} or \citet{Hartmann:1998p664},
\begin{equation}
\Sigg(r) = \Sigma_0 \left(\frac{r}{\rc}\right)^{-\gamma}
\exp\left[\left(-\frac{r}{\rc}\right)^{2-\gamma}\right]
\label{eq:sig_gas}
\end{equation}
(implying the turbulence parameter \alphat is constant), and that the temperature follows a power 
law,
\begin{equation}
T(r) \propto \left(\frac{r}{\rc}\right)^{-q}.
\label{eq:temperature}
\end{equation}
Throughout this article, we assume $\gamma=1$ and $q=1/2$. Although this is not the most general 
case, it does significantly simplify the calculations we will perform, and finds some support 
through the modeling of gas emission lines observed in protoplanetary disks 
\citep[e.g.,][]{Andrews:2012p16676}. We adopt a set of fiducial model parameters: a stellar host
mass of 0.8 M$_{\odot}$, initial disk mass of 0.01 M$_{\odot}$, initial radius ($r_c$) of 20 AU,
\alphat = $10^{-3}$, initial dust-to-gas ratio (\eps) of 0.01 at all $r$, and a (fixed) temperature
scaling such that $T(\mathrm{1\,AU}) = 200$\,K. Here and in Section \ref{sec:late_phase} we will 
keep the gas surface densities fixed for clarity, but we permit \Sigg to evolve viscously in 
Section~\ref{sec:comparison}.

\begin{figure}[t]
\figscale{0.6}
\plotone{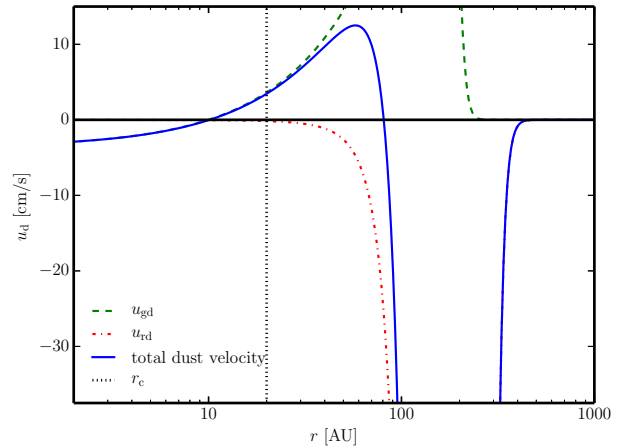}
\figcaption{The early phase radial dust transport velocities for the fiducial model ({\it solid}), 
with the individual contributions from radial drift ({\it dash-dot}) and viscous gas drag ({\it 
dash}) shown separately. The initial \rc is marked by a vertical dotted line. Note that inward 
velocities are negative. \label{fig:vel_example}}
\end{figure}

The initial size distribution of the dust incorporated into a disk is thought to resemble the
interstellar medium \citep[][hereafter MRN]{Mathis:1977p789}, perhaps with some growth up to $\mu$m
sizes \citep[e.g.,][]{Ossenkopf:1994p4887}. A common feature of such size distributions, derived
either theoretically \citep[e.g.,][]{Tanaka:1996p2320,Birnstiel:2011p13845} or observationally
(e.g., MRN), is that most of the mass is contained in the largest grains. Since the transport
velocities noted above also increase with grain size, we assume that the effective dust transport
velocity is approximately equal to that of the largest grains. Here, we assume an initial size of 1
$\mu$m for the dust particles.

With those assumptions and adopted fiducial parameters, Figure \ref{fig:vel_example} shows the
radial velocity distribution of the dust surface density in the early phase, as well as the
decomposed contributions from gas drag and radial drift individually, based on
Eq.~\ref{eq:radial_drift} and \ref{eq:u_gas_drag}, respectively. In the following sections, we
consider in more detail first the effects of drift only (Section 2.1), and then the combined impact
of both drift and drag (Section 2.2) on the overall dust dynamics.

\subsection{Radial drift only}

\citet{Youdin:2002p1262} derived an analytic solution to the advection equation for a case where
radial drift is the only transport mechanism, assuming a fixed particle size and a power-law gas
surface density profile (of infinite extent). They demonstrated that drift creates an outer boundary
in the distribution of mm-sized particles that steepens as it moves in to smaller radii \citep[see
also][]{Jacquet:2012p19670}. Similarly, \citet{Weidenschilling:2003p21295} mentioned the truncation
of the dust disk in models which included the growth and radial drift of dust particles.

Here, we extend the work of \citet{Youdin:2002p1262} by recognizing that this mechanism is more
general, and applies to much smaller particle sizes if the coupling between the gas and dust is of
the right strength. We argue that this is the case in the outer disk, since viscous models have an
exponential taper in \Sigg rather than a fixed power-law profile that extends indefinitely (see
Eq.~\ref{eq:sig_gas}). The corresponding steep decrease in the gas pressure in the outer disk (i.e.,
the large $\mathrm{d}P/\mathrm{d}r$ in Eq.~\ref{eq:radial_drift}) also enhances the dust/gas
coupling (cf., Eq.~\ref{eq:stokes_number}), thereby dramatically boosting the drift velocity of an
initial population of small grains.

For small particles with $\St < 1$, the drift velocity in Eq.~\ref{eq:radial_drift} can be
approximated as 
\begin{equation}
u_\mathrm{rd} \simeq - \,\frac{\St_0\,\cc^2}{\Vc}\,\frac{r}{\rc} \left(\frac{r}{\rc} + \frac{11}{4}\right) \,\exp\left(\frac{r}{\rc}\right),
\label{eq:drift_approx}
\end{equation}
where $\St_0$ is the Stokes number for grains of size $a_0$ at a surface density \Signull, and \cc
and \Vc are the sound speed and Keplerian velocity at \rc, respectively. It is immediately clear in
Eq.~\ref{eq:drift_approx} that the inward speed beyond \rc is exponentially rising, becoming much
faster than for a power-law \Sigg model. This causes even $\mu$m-sized particles to effectively 
spiral inwards, producing a sharp outer edge in the dust surface density distribution and leaving 
the very outer disk devoid of dust.

\begin{figure}[t]
\figscale{0.5}
\plotone{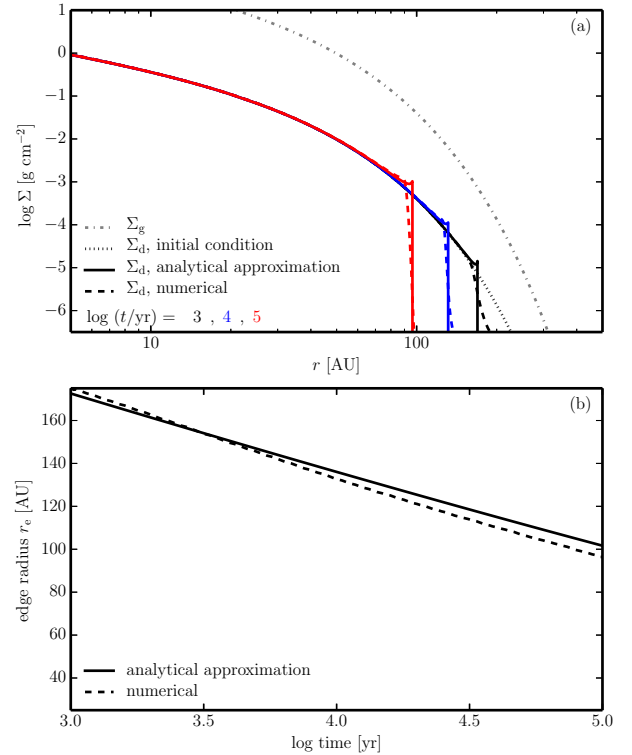}
\figcaption{(a) Analytical ({\it solid}) and numerical ({\it dashed}) calculations of how radial
drift changes the dust surface density profile (cf., Appendix~\ref{app:fixed_size_only_drift}),
leading to the formation of a sharp outer dust edge. The gas surface density profile is shown as a
{\it dotted} curve. (b) The approximate analytical (\textit{solid}) and numerical ({\it dashed})
evolution of the outer dust edge location, $r_\mathrm{e}$ (in units of \rc). 
\label{fig:fixed_a_tapered}}
\end{figure}

After a basic introduction to the relevant dust transport equations (Appendix~\ref{app:general}), we
derive an analytic solution in Appendix~\ref{app:fixed_size_only_drift} that describes the early
evolution of the dust surface densities, $\Sigd(r,t)$ (cf., Eq.~\ref{eq:sigd_fixed_size_drift}), and
the location of this outer dust edge, $r_\mathrm{e}(t)$ (cf., Eq.~\ref{eq:redge_fixed_size_drift}).
Figure~\ref{fig:fixed_a_tapered} compares these analytic solutions (Eqs.~\ref{eq:fixed_size_time},
\ref{eq:sigd_fixed_size_drift}, and \ref{eq:redge_fixed_size_drift}) to a direct, numerical
calculation for a single particle size ($a_0 = 1~\mu$m). The top panel shows discrete steps in
$\Sigd(r,t)$, and demonstrates that small dust is quickly removed beyond a sharp outer edge that
drifts inward with time. The bottom panel tracks that edge location directly, finding that
$r_\mathrm{e}(t)$ decreases from $\sim$9 to 5~\rc (in this example, 175 to 100 AU). As a comparison,
we evolved the same initial dust distribution for a model with a power-law gas surface density
profile, $\Sigg = \Sigma_0 (r/\rc)^{-1}$ (not shown in Fig.~\ref{fig:fixed_a_tapered}), and
confirmed that \Sigd does not appreciably change over the $\sim$0.1\,Myr timescale on which
these calculations are valid.  Figure~\ref{fig:fixed_a_tapered} shows slight differences in the edge 
position between the approximate analytical and numerical solutions: these are entirely due to the 
approximation used for the dust velocity, which becomes better for smaller \rc. Solving 
Eqns.~\ref{eq:general_solution} and \ref{eq:t_integral} using the exact velocity term (as is 
necessary in the next section) significantly improves their agreement.

It is worth a reminder that these solutions have so far neglected particle size evolution, by 
assuming a grain size that does not evolve with time.  Ultimately, grain growth limits the 
applicability of our assumptions for this early phase, and effectively bounds the inward motion 
of the outer dust edge.  That limit can be estimated numerically by evaluating 
Eq.~\ref{eq:redge_fixed_size_drift} at the growth timescale, $r_\mathrm{e}^{\mathrm{min}} \simeq 
r_e(\tau_\mathrm{g})$.  For our fiducial model, $r_\mathrm{e}^{\mathrm{min}}$ is roughly 6~\rc, and 
for reasonable ranges of the key input parameters, we suggest that $r_\mathrm{e}^{\mathrm{min}}$ can 
range between $\sim$3 and 8~\rc.  At very large distances outside this edge, the gas densities 
become so low that even small grains have $\St \gg 1$, and therefore the adopted approximations for 
dust transport in this phase do not apply (at large \St, drift velocities are lower than assumed).  
In our fiducial model, this occurs for $r \gtrsim 14~\rc$, although it should be noted that the dust 
surface densities at such radii are negligibly small.  To summarize, the grain growth timescale 
imposes limits on the applicability of the early phase that can be translated into a specific radial 
range in the disk; in the case of our fiducial model, that range is $\sim$120--280\,AU.

\subsection{Radial drift and gas drag}

Although we showed in Figure \ref{fig:vel_example} that radial drift dominates the dust motion in
the outer disk during the early phase, it is not the only relevant transport mechanism. Dust can
also be dragged along with the viscously evolving gas disk. For $\St < 1$, this radial motion is
approximately equal to Eq.~\ref{eq:u_gas}. We can get some intuition on how the dust evolves if both
radial drift and gas drag are considered together by assuming the same, fixed \Sigg as in
Eq.~\ref{eq:sig_gas} and solving for the dust transport equations in Appendix~\ref{app:general}
using the sum of Eq.~\ref{eq:radial_drift} and Eq.~\ref{eq:u_gas_drag} as the effective velocity
term. This assumption neglects the effects of evolution in \Sigg, which would slightly modify the
velocity term, but since significant surface density changes occur over the relatively long viscous
timescale in the outer disk, these modifications have only a minor effect on the results during the
early phase.

\begin{figure}[th]
\figscale{0.45}
\plotone{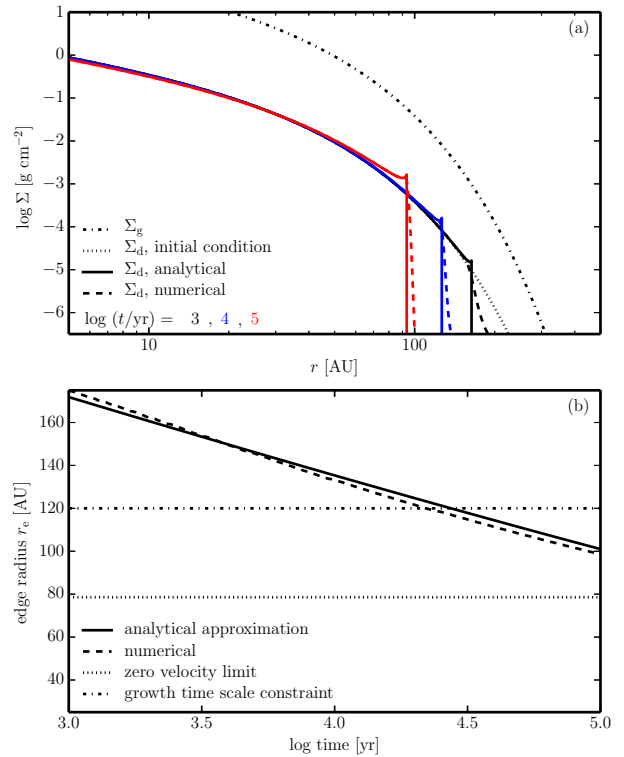}
\figcaption{(a) Analytical ({\it solid}) and numerical ({\it dashed}) calculations of the dust
surface density evolution in the early phase as in Fig.~\ref{fig:fixed_a_tapered}, but in this case
including gas drag. Note the enhancement of \Sigd near the dust edge, produced by the outward motion
of dust entrained in the viscous gas flow. (b) The evolution of the outer dust edge including both
radial drift and gas drag. In specific cases where the growth timescales (i.e., the duration of the
early phase) are long or \alphat is high, the inward motion of the dust edge will be halted at a
limiting radius where drift and drag are balanced (close to 80~AU in this case, see dotted line).
Otherwise, the ultimate location of the dust edge will be set by the grain growth timescale
($\sim$120~AU for our parameters, cf. dash-dotted line).
\label{fig:drift_drag}}
\end{figure}

The impact on $\Sigd(r,t)$ and $r_\mathrm{e}(t)$ when gas drag and radial drift are incorporated
together is described analytically in Appendix~\ref{app:fixed_size_and_gasdrag} and shown
graphically in Figure~\ref{fig:drift_drag}. The outermost regions behave similarly to the drift-only
case, since the $u_\mathrm{rd}$ term dominates at large $r$. However, just inside the dust edge
there is a notable increase in \Sigd due to the outward motion of dust entrained in the viscous gas
flow. Figure \ref{fig:vel_example} demonstrates that this gas drag effect dominates the dust
transport in a relatively narrow radial range. In essence, the \textit{outward}-directed
$u_\mathrm{gd}$ from the viscously expanding gas disk and the \textit{inward}-directed
$u_\mathrm{rd}$ converge near $r_\mathrm{e}$ and lead to a pile-up of dust.

The location of $r_\mathrm{e}$ shifts inward with time as in the drift-only case, but at some point
the outward gas drag dominates and limits this evolution. This implies that the dust edge
asymptotically approaches a minimum radius ($r_\mathrm{e} \ge r_\mathrm{lim}$) that can be 
estimated analytically,
\begin{equation}
\frac{r_\mathrm{lim}}{\rc} =
\mathcal{W}\left[\frac{3\,\alphat}{\St_0}\,\exp\left(\frac{11}{4}\right)\right]-\frac{11}{4},
\label{eq:r_lim}
\end{equation}
where we have used the Lambert $\mathcal{W}[\cdot]$ function (see
Appendix~\ref{app:fixed_size_and_gasdrag}). The location of $r_\mathrm{lim}$, roughly 4~\rc (80 AU)
in our fiducial model, is marked as a dotted line in the bottom panel of
Figure~\ref{fig:drift_drag}. For our fiducial model, $r_\mathrm{lim}$ is well inside the limit to 
the inward motion of the dust edge imposed by the grain growth timescale constraint described above 
($r_\mathrm{e}^\mathrm{min}$). But in principle, this drag-imposed limit on the dust edge location 
could be relevant in disks that are more turbulent (larger \alphat) or have intrinsically longer 
growth timescales (e.g., brown dwarf disks).  

To summarize, we have demonstrated analytically and numerically that the observational features
noted in Section 1 --- a sharp outer dust edge and corresponding decrease in the dust-to-gas
ratio --- are naturally generated by radial drift and, to a lesser extent gas drag, during an early
phase in disk evolution, before substantial particle growth or viscous evolution have developed. In
our fiducial example, the dust surface densities have decreased dramatically outside $\sim$5~\rc
($\sim$100~AU) within 0.1 Myr, before the grains have had a chance to grow beyond a few $\mu$m in
size.

\section{Late Phase}\label{sec:late_phase}

While the ``early phase" discussed in the previous section represents a good approximation for dust
transport in the outer disk, the dust in the inner disk is evolving very quickly and cannot be
treated as if it has a fixed grain size. The particle size distribution and its associated
velocities are time-dependent and spatially varying at smaller radii, and therefore the system is
best studied with detailed numerical simulations (see Section~\ref{sec:comparison}). However, if
particle growth has progressed to the point that it is limited by effects like fragmentation or the
rapid removal of larger grains due to radial drift, we can define an analogous ``late phase," where
particles have reached a maximum equilibrium size that depends on their distance to the host star.
The timeframe where this late phase is valid depends intimately on the initial conditions and other
parameters involved in the model, but a rough estimate of a few $\times$ $10^5$~years is a
reasonable approximation.

\subsection{Drift limited case}\label{sec:late_phase:drift}

\citet{Birnstiel:2012p17135} showed that, if particles are not first subject to a barrier to their
growth, the motion induced by radial drift itself imposes a size limit by removing large particles
faster than collisional growth can replenish them. That limit is set to the size where the
corresponding drift and growth timescales are roughly equal, and has an associated drift velocity
\begin{equation}
u_\mathrm{rd} = - f_\mathrm{d} \,S\, \eps \,\Vk,
\label{eq:u_dust_driftlim}
\end{equation}
where the order-of-unity constant $f_\mathrm{d}$ was taken to be $0.3$. Comparing the drift
limited transport velocity of Eq.~\ref{eq:u_dust_driftlim} with the gas drag velocity, it can be
shown that $u_\mathrm{rd} > u_\mathrm{gd}$ so long as $\eps \gtrsim 0.01\alphat$.

\begin{figure}[t]
\figscale{0.45}
\plotone{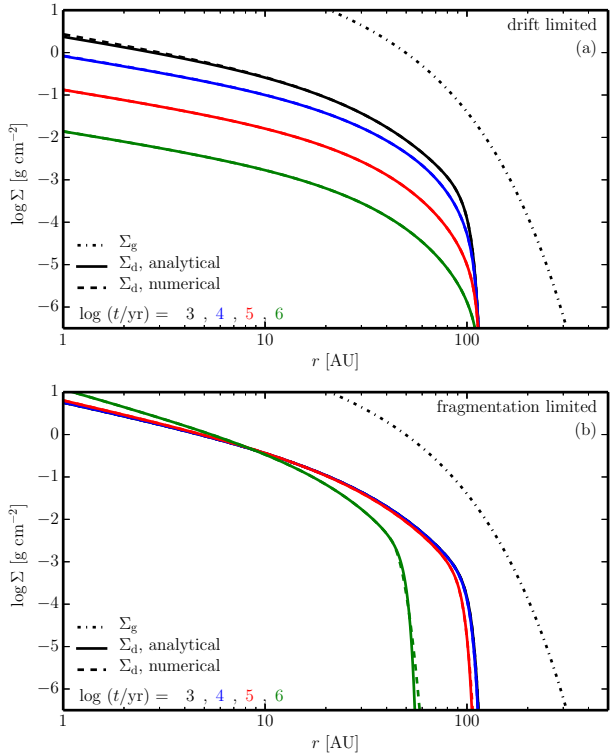}
\figcaption{The subsequent evolution of the dust surface density distribution during the ``late 
phase," when the maximum particle size is limited either by removal due to radial drift (a) or by 
grain fragmenting collisions (b). The notation is the same as in Figures~\ref{fig:fixed_a_tapered} 
and \ref{fig:drift_drag}.
\label{fig:frag_and_growth}}
\end{figure}

We derive an analytic solution for $\Sigd(r,t)$ in Appendix~\ref{app:drift_limited}, assuming that
dust is transported with this drift limited velocity. The resulting behavior is shown in 
Figure~\ref{fig:frag_and_growth}(a), along with a comparison to the corresponding direct
numerical calculations. Unlike in the early phase described in Section~\ref{sec:early_phase}, drift
according to Eq.~\ref{eq:u_dust_driftlim} will not form a sharp outer edge by itself. Therefore, in
this example, we have imposed an initial condition informed by the early phase results, such that
the dust-to-gas ratio is constant in the inner disk ($\eps = 10^{-2}$), but drops steeply beyond
$r_\mathrm{e} \simeq 5~\rc$. Figure~\ref{fig:frag_and_growth}(a) shows that the dust disk will
effectively retain this shape: the location of the edge is fixed, but the drop-off in \Sigd gets
progressively smoother as the dust mass homologously drains inward. The dust surface density in the
inner disk is fed by the inward-drifting dust from the outer regions, and therefore follows the
steady-state profile derived by \citet{Birnstiel:2012p17135}: the radial power-law index of \Sigd 
(where $\Sigma \propto r^{-p}$) is $p_\mathrm{d} = (2\,p_\mathrm{g}+1)/4$, which for a local gas 
surface density index of $p_\mathrm{g} \simeq 1.2$ gives rise to $p_\mathrm{d} \simeq 0.85$, in 
agreement with the measured slope of 0.90 in Figure~\ref{fig:comparison}.\footnote{Note that 
$p_\mathrm{g} > \gamma$ due to the slight, but non-negligible, influence of the exponential 
tapering {\it inside} of \rc.}

Transport along this radial drift barrier is self-regulating in the sense that the sizes particles
can reach, and therefore also their drift speeds, depend on the dust-to-gas ratio, \eps.
Consequently, a high \eps produces fast drift motion and a rapid depletion of dust mass in the 
outer disk. Once \eps is reduced as the disk evolves, particles are smaller and the decay of \Sigd
proceeds more slowly. As can be seen from Eq.~\ref{eq:u_dust_driftlim}, the sticking efficiency,
$S$, acts in the same way: if $S$ were 0.5 for the relevant grain sizes, then the drift timescales
are twice as long, and the maximum sizes are twice as small.

\subsection{Fragmentation limited case}\label{sec:late_phase:frag}

A solution to the transport equations for the case where fragmentation limits the maximum grain
sizes is derived in Appendix~\ref{app:frag_limited}. As for the drift limited case, no sharp outer
edge is formed by this scenario itself. So, again, we impose an outer edge in the same way as the
previous section, with an initial condition on $\eps(r)$ based on the calculations in
Section~\ref{sec:early_phase}. As long as there is still a significant outer disk that transports
dust inwards, the inner regions in this scenario will generate a stationary \Sigd profile with a
power-law index $p_\mathrm{d} \simeq 1.5$, as derived by \citet{Birnstiel:2012p17135}. 

The evolution of \Sigd in the fragmentation limited case, shown in
Figure~\ref{fig:frag_and_growth}(b), is not self-regulating: particles drift at a constant rate 
that can be either fast or slow, depending on their size. Consequently, the outer dust edge moves 
inward as the disk evolves. If grain growth is limited by fragmentation, the edge can therefore be 
located substantially closer to the central stellar host.

\section{Comparison to detailed simulations}\label{sec:comparison}

We described above how dust transport processes in the ``early" and ``late" phases affect the dust
surface density evolution in an analytic framework. However, to do this we had to make some
important simplifications. In this section, we will compare this framework to a full suite of
numerical simulations to test how well it describes the complicated behavior of a more realistic
protoplanetary dust disk, where the constituent particles are growing and fragmenting upon
collisions, migrating inwards due to radial drift, and at the same time being mixed and dragged
along with the turbulent gas reservoir in which they are embedded. These detailed numerical
simulations act as a reality check to test how well the adopted analytic simplifications hold, or
whether neglected effects (e.g., particle size evolution, turbulent mixing) can cause significant
deviations from the analytic results. The code used for these simulations was originally presented
by \citet{Birnstiel:2010p9709}. As initial conditions, we again assume a 0.8~$M_\odot$ stellar host,
and adopt a \Sigg profile following Eq.~\ref{eq:sig_gas}, with an initial $\rc = 20$~AU, $\eps =
0.01$ (at all $r$), $\alphat=10^{-3}$, and disk mass of $0.01~M_\odot$. The initial particle size
distribution was assumed to have a MRN-like slope, $n(a) \propto a^{-3.5}$, between 0.1 and
1~$\mu$m.

\begin{figure*}[t!]
\figscale{1}
\plotone{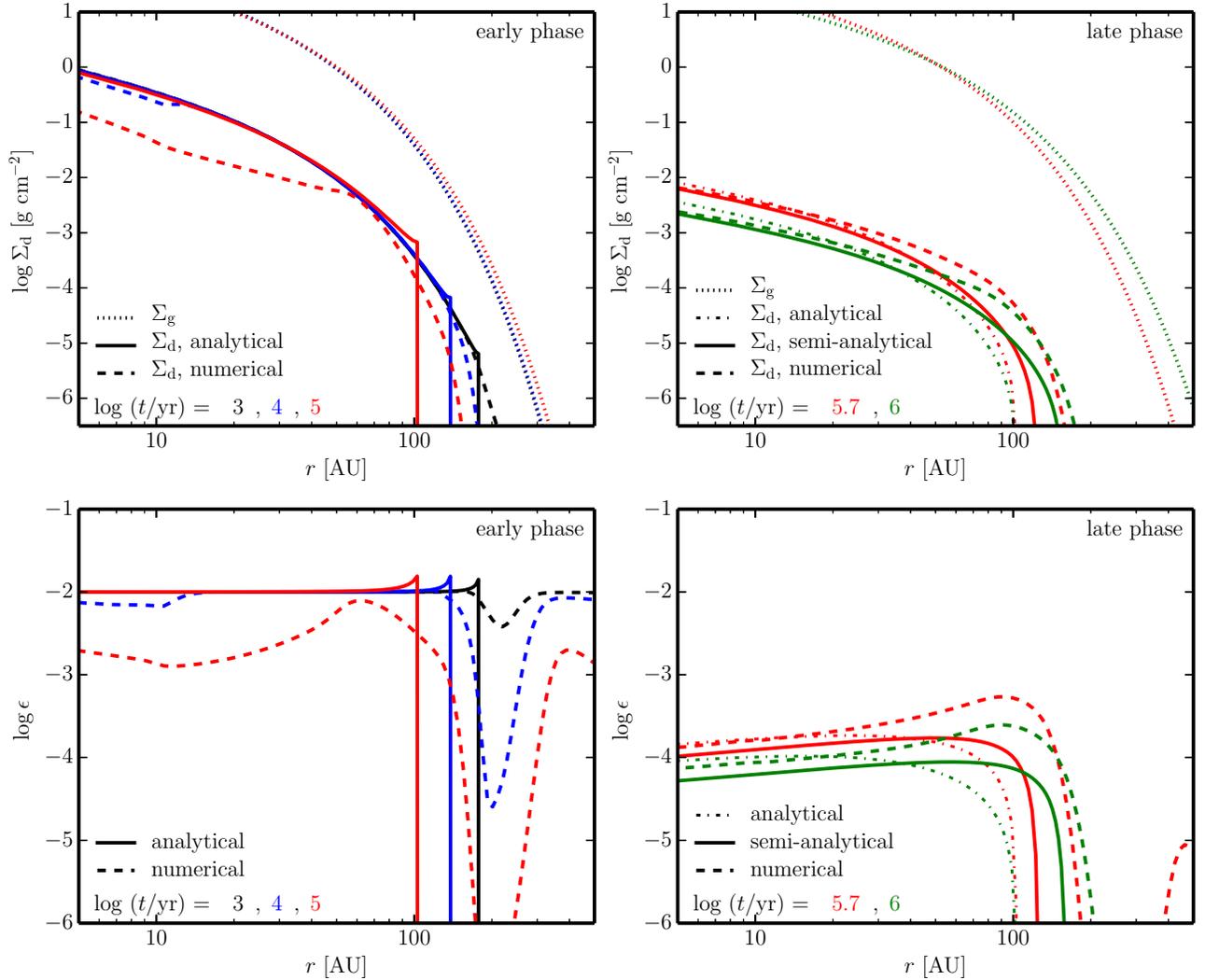}
\figcaption{A comparison between the predictions of the analytic framework (\textit{solid} and
\textit{dash-dotted}) and the detailed numerical simulations (\textit{dashed}) for the evolution of
\Sigd ({\it top}) and \eps ({\it bottom}) during the ``early" and ``late" phases ({\it left} and
{\it right}, respectively). The snapshots correspond to $10^3$, $10^4$, and $10^5$ years in the 
early phase and to $5\times 10^5$, and $10^6$ years in the late phase. The simulations confirm that 
a sharp outer dust edge is formed quickly in both the analytic approximation and the numerical 
simulations, with deviations primarily related to the added sophistication of including particle 
growth and an initial distribution of particle sizes. The late phase in both cases are remarkably 
similar, considering the simplicity of the analytic approximation. \label{fig:comparison}}
\end{figure*}

Figure \ref{fig:comparison} shows representative discrete steps in the evolutionary behavior of
$\Sigd(r,t)$ and $\eps(r,t)$ for the full numerical simulation, broken down into approximations of
the ``early" (\textit{left}; $\lesssim$0.1\,Myr) and ``late" (\textit{right}; $\gtrsim$0.1\,Myr)
phases to facilitate a direct comparison with the analytic prescriptions described in Sections
\ref{sec:early_phase} and \ref{sec:late_phase}, respectively, for the same initial conditions. 
Recall the discussion in Section 2.1 that showed how the analytical calculations for the early phase 
are only strictly valid for the radial range $\sim$120--280\,AU. In the numerical simulations of the 
early phase, the dust-to-gas ratio in the inner disk deviates by about an order of magnitude from 
the analytic solution due to the onset of grain growth, an effect that was explicitly not treated in 
the simplified derivations of Section~\ref{sec:early_phase} 
(Appendix~\ref{app:fixed_size_and_gasdrag}). The numerical simulations also quickly form a sharp
outer edge in the dust distribution at a similar location as predicted by the simple analytic
solution, albeit with some differences in the details. The simulated edge is less sharp than the
analytic prediction, partly due to real diffusion (because turbulent mixing is included in the
simulations) but also due to numerical diffusion.\footnote{In the previous sections, we used higher
order advection schemes to demonstrate the agreement between the analytic and direct numerical
solutions in the simplified framework. But in this section, the numerical advection of the dust
evolution code is an implicit donor-cell scheme which is intrinsically more diffusive. At the same
time, the radial resolution is necessarily lower due to the high computational costs of running the
full dust evolution model.} Moreover, unlike the analytic solutions, the numerical code does not
assume a monodisperse size distribution, but rather uses an MRN-like distribution, as an initial
condition. Since each particle size is associated with a distinct outer edge location (cf.,
Eq.~\ref{eq:r_lim}), a size distribution naturally produces a smoother edge. In this example,
$\mu$m-sized grains have drifted further inward (to $\sim$100~AU) than the 0.1~$\mu$m-sized grains
(which extend beyond 150~AU).  Beyond $\sim$280\,AU, the growth time scales again limit the 
applicability of the early phase solution (cf., Sect.~2.1), which explains the discrepancies at 
those distances. Despite these differences, it is clear that the fundamental prediction of the 
analytic models -- that a relatively sharp edge in the dust distribution is formed quickly and 
maintained on longer timescales -- is reproduced with good fidelity by the more sophisticated 
numerical simulations.

The right-hand panels of Figure~\ref{fig:comparison} show the continued evolution of the disk in the
``late" phase, where we have evolved the analytic result according to the drift limited scenario
described in Appendix~\ref{app:drift_limited} (cf., dash-dotted curves). There are differences in 
the outer edge location in this comparison, stemming from the fact that the analytical drift 
solution does not include the viscous evolution of the gas disk. To remedy that, we included this 
effect in a semi-analytical way (see Appendix~\ref{app:coadvection}): the results are shown in the 
right panels of Figure~\ref{fig:comparison} as solid curves. There is remarkable agreement between 
the shape and overall absolute values of the surface densities between the numerical simulations and 
the semi-analytic results, especially considering the amount of simplification used to develop the 
latter. The semi-analytic results under-predict \Sigd in the outermost regions of the disk, partly 
because we neglected diffusion in that scenario, but mainly because the simple assumption that 
turbulence drives the collisional (and therefore growth) timescales (i.e., Eq.~\ref{eq:tau_grow}) 
becomes less accurate in the outer disk: at such low densities, drift-induced relative velocities 
become comparatively important \citep{Birnstiel:2012p17135}. In addition, our (semi-)analytical 
models used a mass-averaged size of 0.47\,$\mu$m instead of the fixed assumption of 1\,$\mu$m 
adopted in Section 2.1. A mass-averaged size is a good approximation in these calculations 
initially, however at later times the outer disk in the numerical simulation contains mostly smaller 
grains. This is why the viscous spreading of the outer edge proceeds further than in the analytical 
curves in Figure 5.


\section{Discussion}\label{sec:conclusion}

We have developed an analytic framework that describes the bulk transport of solids embedded in a
gas-rich, viscous accretion disk during two simplified evolutionary epochs. During an ``early" 
phase in that evolution, before dust grains in the outer disk have had a chance to grow
($\lesssim$0.1~Myr), we find that the exponential taper in the gas surface densities beyond a
characteristic radius (\rc) substantially boosts the radial drift velocities for particles {\it of
any size}. That inward migration naturally produces a sharp edge to the dust density distribution,
or equivalently a steep drop in the dust-to-gas mass ratio, in the outer disk, representing a
distinct and potentially observable ``fingerprint" of the drift process. The location of this edge
shifts inward with time, to a limit imposed by the grain growth timescale or viscous drag in the 
outward-spreading gas disk. While that location depends on detailed model parameters (particularly 
the grain size, gas surface density, and turbulence parameter), we expect typical edge radii of 
$\sim$3--8$\times$ the initial \rc (roughly 30--300~AU).

During a ``late" phase of this evolution, once solid particles have grown larger (perhaps a few
$\times10^5$~yr), we solved the transport equations under the assumption that particle sizes are
limited either by the drift or fragmentation barriers \citep[cf.,][]{Birnstiel:2012p17135}.
Although the latter might be relevant initially, the decreasing dust-to-gas ratio will eventually
ensure that drift is the barrier that limits particle growth in the outer disk. With this analytic
prescription, we showed that the edge formed during the ``early" phase is preserved, and its
location does not change significantly (if the gas profile is fixed) or moves slightly outward
(for a viscously spreading gas disk). A direct comparison to more sophisticated, time-dependent
numerical simulations of the growth and transport of disk solids demonstrated that our simplified
analytic framework faithfully captures the important features of the dust distribution in both the
early and late phases.

That said, we should emphasize that this analysis does nothing to {\it solve} the ``radial drift
problem" (frequently termed the ``m-size barrier", although ``mm/cm-size barrier" is more
appropriate for the outer disk): there are still fundamental issues with the derived evolution
timescales that are generally inconsistent with observations
\citep[e.g.,][]{Dominik:2007p1420,Brauer:2007p232}. One promising solution relies on pressure traps
to stop or slow the radial drift of dust grains in the outer disk, generated either by turbulence 
or interactions with an embedded companion \citep[e.g.,][]{Pinilla:2012p16999,Pinilla:2012p18741}.
Note that traps of the right strength ($\mathrm{dln}{P}/\mathrm{dln}r \ge 0$) may present an 
alternative means of creating an outer edge in the dust distribution, for the same physical reasons 
outlined here.

The most fundamental (and perhaps obvious) conclusion from this study is that dust transport
processes in a viscous accretion disk quickly modify the radial distribution of dust-to-gas ratios,
$\eps(r)$, in the outer disk. Taking these calculations at face value, we must conclude that the
typical assumption of $\Sigd \propto \Sigg$ in a protoplanetary disk is internally inconsistent.
Therefore, the common practice of inferring \Sigd from observations based on a parametric
prescription for a viscous gas disk \citep[like Eq.~\ref{eq:sig_gas};
e.g.,][]{Andrews:2009p7729,Andrews:2010p17519,Isella:2009p7470,Guilloteau:2011p15287} is not
supported by any physical motivation. Instead, examination of the typical dust distributions 
derived from solving the transport equations suggests that a more reasonable parametric 
approximation for $\Sigd(r)$ would be a multi-stage power-law with indices determined by the dust 
physics: $p_\mathrm{d} \simeq 1.5$ for a fragmentation-limited dust distribution in the inner disk, 
a smaller $p_\mathrm{d}$ for a drift-dominated dust distribution in the outer disk, and then a 
steep drop (large $p_\mathrm{d}$) outside of $r_\mathrm{e}$.

The key results from our dust transport calculations regarding the structural distributions of gas
and solids in protoplanetary disks find strong, qualitative support from recent observations. A
growing sample of disks have exhibited evidence for a {\it size} discrepancy in comparisons of their
mm-wave emission in the CO line and associated continuum: the gas always appears more spatially
extended than the dust
\citep{Panic:2009p11789,Andrews:2012p16676,deGregorioMonsalvo:2013p21848,Rosenfeld:2013p21297}.
Radiative transfer models confirm that this discrepancy is not an artifact of limited sensitivity,
or optical depth effects. The dust transport calculations described here naturally reproduce this
size discrepancy, due to the steep decrease in $\eps(r)$ near $r_\mathrm{e}$ created during the
``early" phase of evolution. It is worth a reminder that, in the ``late" phase, gas drag in the
outward viscous flow can dominate drift velocities for small ($\mu$m-sized) particles that were
originally confined inside $r_\mathrm{e}$ (indeed, this effect contributes to the smearing of the
edge feature in the full numerical simulations presented in Section~\ref{sec:comparison}). This
size-sorting offers a potential explanation for the spatial segregation of mm and $\mu$m-sized
grains, where the latter appear in scattered light images to be more extended (e.g., compare
\citealp{Weinberger:2002p21098} and \citealp{Andrews:2012p16676}). Finally, very high quality
mm-wave continuum emission measurements are starting to show the definitive signature of the {\it
sharp} outer edge in the dust distribution that we have advocated, in the form of a distinctive
oscillation pattern in their interferometric visibilities
\citep{Andrews:2012p16676,deGregorioMonsalvo:2013p21848}.

The emerging consensus between observations and the theoretical framework developed here makes a
strong case that dust transport dominated by radial drift plays a significant role in shaping the
spatial distribution of disk solids. Ultimately, sensitive, high resolution mm-wave datasets should
be able to place more quantitative constraints on this transport process. If observers continue to
find sharp dust continuum edges and gas/dust size discrepancies in nearby disks, the combination of
such measurements could in principle be linked to the key parameters $r_\mathrm{e}$ and \rc, and
thereby indirectly to $\eps(r)$ (which would help guide any revisions to transport simulations).
Making a direct link between our simulations and observations is unfortunately complicated by the 
radial size-sorting that is also naturally imprinted by grain growth. While the {\it total} dust 
mass is confined within $r_\mathrm{e}$, the larger grains traced by radio interferometers are 
expected to be preferentially located at even smaller radii. In practice, some further development 
of our calculations and a link with radiative transfer models will be required to quantitatively aid 
the interpretation of mm/cm-wave continuum emission. With the start of science operations with the 
Atacama Large Millimeter/submillimeter Array (ALMA) and the recent upgrade of the Karl G.~Jansky 
Very Large Array (VLA), the prospects are excellent for continued progress on this topic.

\acknowledgments
We are grateful for support from the NASA Origins of Solar Systems grant NNX12AJ04G, computing time
on the Smithsonian Institution high performance cluster, {\tt hydra}, and for a thoughtful review by
the anonymous referee which helped improve the quality of the article.
\bibliographystyle{aa}
\bibliography{/Users/til/Documents/Papers/bibliography}
\clearpage
\appendix

\section{A. General Solution}\label{app:general}
The general equation that describes the advective radial transport of the dust surface density in
cylindrical coordinates is
\begin{equation}
\frac{\partial \Sigd}{\partial t} +\frac{1}{r}\frac{\partial}{\partial r}\left(r\,u(r)\,\Sigd\right)
= 0.
\label{eq:advection}
\end{equation}
Since the prescription we adopt to describe the gas surface densities is written in terms of the 
radius \rc, it is convenient to introduce the dimensionless variable $x = r/\rc$, such that 
\begin{equation}
\frac{\partial \Sigd}{\partial t} +\frac{1}{x}\frac{\partial}{\partial
x}\left(\frac{x\,u(x)}{\rc}\,\Sigd\right) = 0.
\label{eq:A2}
\end{equation}
The characteristic equations for this partial differential equation read
\begin{eqnarray}
\frac{\mathrm{d}t}{\mathrm{d}s} &=& 1\\
\frac{\mathrm{d}x}{\mathrm{d}s} &=& \frac{u(x)}{\rc}\\
\frac{\mathrm{d}\Sigma(s)}{\mathrm{d}s} &=& - \frac{\Sigma(s)}{\rc\,x} \, \frac{\partial
\left(u(x)\,x\right)}{\partial x}.
\end{eqnarray}
The solution to Eq.~\ref{eq:A2} then can be written as
\begin{equation}
\Sigma(x,t) = \Sigma(x_0,0)\, \frac{u(x_0)\,x_0}{u(x)\,x},
\label{eq:general_solution}
\end{equation}
where $x_0$ is the position of the characteristic $s$ at dimensionless radius $x$ at time $t=0$, 
which is indirectly defined as
\begin{equation}
\frac{t}{\rc} = \int_{x_0}^{x}\,\frac{1}{u(x')}\,\mathrm{d}x'.
\label{eq:t_integral}
\end{equation}
If the integral in Eq.~\ref{eq:t_integral} can be carried out and solved for $x_0$, the result 
represents the full, analytical solution to the advection equation. But even if this cannot be 
carried out analytically, a numerical evaluation of Eq.~\ref{eq:t_integral} is usually much faster 
than discretizing Eq.~\ref{eq:advection} and evolving it iteratively.

\section{B. Radial drift for a fixed particle size and $p_\mathrm{gas} =
1$}\label{app:fixed_size_only_drift} In the following, we assume a time-independent gas surface density
\begin{equation}
\Sigg(r) = \Signull \, \left(\frac{r}{\rc}\right)^{-p_\mathrm{gas}}\exp\left[-\left(\frac{r}{\rc}\right)^{2-p_\mathrm{gas}}\right],
\label{eq:sigg}
\end{equation}
where $p_\mathrm{gas}=1$ (as in Eq.~\ref{eq:sig_gas}) and a radial temperature dependence of 
\begin{equation}
T(r) = \Tc \left(\frac{r}{\rc}\right)^{-q},
\end{equation}
as in Eq.~\ref{eq:sig_gas} and \ref{eq:temperature}, respectively. In this structural 
parameterization and for a (radially) constant grain size without any growth, the dust drift 
velocity can be written as 
\begin{equation}
u = -A
\exp\left(x\right)\,x^{-q+3/2}\,\left(x+\frac{q+5}{2}\right),
\end{equation}
where
\begin{equation}
A = \frac{\St_0\,\cc^2}{\Vc}
\end{equation}
(see Section~\ref{sec:late_phase:drift} for symbol definitions). The system can be solved by 
numerically evaluating Eq.~\ref{eq:t_integral}, however we can find a good approximation for large 
$x$ (the outer disk) using the velocity
\begin{equation}
u = -A
\exp\left(x\right)\,x^d,
\end{equation}
where $d = \frac{5}{2}-q$. With that approximation, Eq.~\ref{eq:t_integral} can be written
\begin{equation}
t(x,x_0) =
-\frac{\rc}{A}\,\left(\Gamma\left[1-d,x_0\right]-\Gamma\left[1-d,x\right]\right),
\label{eq:fixed_size_time}
\end{equation}
where $\Gamma(a,z)=\int_z^{\infty} t^{a-1} e^{-t} \,\mathrm{d}t$ is the incomplete gamma function.
The approximate solution for the dust surface density evolution is therefore
\begin{equation}
\Sigma(x,t) = \Sigma(x_0,0) \, \exp \left[-\left(x-x_0\right)\right] \,
\left(\frac{x}{x_0}\right)^{-d-1}.
\label{eq:sigd_fixed_size_drift}
\end{equation}

For $x\gg 1$, we can replace the gamma function in Eq.~\ref{eq:fixed_size_time} with $\Gamma(1-d,x) 
\simeq \exp(-x)\,x^{-d}$ and solve for $x_0$ to find
\begin{equation}
x_0 = d \cdot
\mathcal{W}\left[\frac{1}{d}\,\left(\mathrm{e}^{-x}\,x^{-d}-\frac{A\,t}{\rc}\right)^{-1/d}\right],
\label{eq:x0_approx}
\end{equation}
where $\mathcal{W}\left[\cdot\right]$ is the Lambert $\mathcal{W}$-function, also known as product
logarithm or omega function (note that the $\mathcal{W}$ function is readily available in most 
mathematical software\footnote{E.g., \texttt{scipy.special.lambertw} in Scipy, \texttt{lambertw} in 
Matlab, \texttt{ProductLog} in Mathematica. However, to our knowledge, this has not yet been 
implemented in IDL.}).

Eq.~\ref{eq:sigd_fixed_size_drift} together with Eq.~\ref{eq:x0_approx} provides a good
approximate solution to the transport equation. There exists a location $x_\mathrm{e}$ where $x_0$ 
diverges, meaning that all $x_0$ outside $x_\mathrm{e}$ have moved inward of this point (i.e., 
$x_\mathrm{e}$ represents the outer edge of the dust disk). This edge location is approximately
\begin{equation}
x_\mathrm{e} = \frac{r_\mathrm{e}}{\rc} = d\cdot
\mathcal{W}\left[\frac{1}{d}\left(\frac{A\,t}{\rc}\right)^{-1/d}\right].
\label{eq:redge_fixed_size_drift}
\end{equation}

\section{C. Including gas drag}\label{app:fixed_size_and_gasdrag}
If we include the gas drag velocity, Eq.~\ref{eq:u_gas_drag}, for $p_\mathrm{gas}=1$ and 
$q=\frac{1}{2}$, we get a velocity of
\begin{equation}
u = B \,\left(x-\frac{1}{2}\right)-A\,x\,\exp(x)\left(x+\frac{11}{4}\right),
\end{equation}
where $B=3\alphat\,c_\mathrm{c}^2/\Vc$. The solution to the advection equation is then given by
\begin{equation}
\Sigd(x,t) = \Sigd(x_0,0)
\frac{x_0}{x}
\frac
    {x_0-\frac{1}{2}-\mathrm{Pe}_0\,x_0\,\exp (x_0)\,\left(x_0+11/4\right)}
    {x  -\frac{1}{2}-\mathrm{Pe}_0\,x  \,\exp (x  )\,\left(x  +11/4\right)},
    \label{eq:fixed_size_and_gasdrag}
\end{equation}
where $\mathrm{Pe}_0=\mathrm{\St_0}/3\alphat$ ($\simeq$ the P{\'e}clet number) and $x_0$ is
defined by
\begin{equation}
t = \rc \int_{x_0}^{x}
\left(B\,\left(x'-\frac{1}{2}\right)-A\,x'\,\exp (x')
\,\left(x'+\frac{11}{4}\right)\right)^{-1}\mathrm{d}x'.
\end{equation}

\section{D. Drift-limited Solution}\label{app:drift_limited}
The velocity of particles in the drift-limited scenario \citep[see][]{Birnstiel:2012p17135} is 
given by
\begin{equation}
u = - f_\mathrm{d}\,S\,\epsilon\,\Vk,
\end{equation}
where $\eps = \Sigd/\Sigg$, implying that, in contrast to above, the velocity is now a function of 
both position and \Sigd. Analogous to the derivation in Appendix~\ref{app:general}, the 
characteristic equations become
\begin{eqnarray}
s &=& f_\mathrm{d}\,S\,t\\
\frac{\mathrm{d} r}{\mathrm{d} s} &=& -2 \, \epsilon\,\Vk\\
\frac{\mathrm{d}\Sigd}{\mathrm{d}s} &=&
\frac{\Sigd^2}{\Sigg}\,\Ok+\Sigd^2\,\frac{\mathrm{d}}{\mathrm{d} r} \left(\frac{\Vk}{\Sigg}\right)
\end{eqnarray}
and yield the solution
\begin{equation}
\Sigd(x,t) = \Sigd(x_0,0) \,
\left(\frac{x}{x_0}\right)^{-1/4}\,\left(\frac{\Sigg(x)}{\Sigg(x_0)}\right)^{1/2},
\label{eq:sigma_drift}
\end{equation}
where $x_0$ is again the initial position of a test particle, such that
\begin{equation}
t = - \frac{\rc \, \sqrt{\Sigg(x_0)}}{2 f_\mathrm{d}\,S\,\Vc \,x_0^{1/4}\,
\Sigd(x_0,0)} \, \int_{x_0}^{x} {x'}^{3/4} \, \Sigg(x')^{1/2} \, \mathrm{d}x'.
\end{equation}

For the gas surface density profile as in Eq.~\ref{eq:sigg} with $p_\mathrm{gas}=1$ or also for 
a slightly more general case of 
\begin{equation}
\Sigg(r) = \Signull \,
\left(\frac{r}{\rc}\right)^{-p_\mathrm{gas}}\exp\left[-\frac{r}{\rc}\right],
\end{equation}
and $p_\mathrm{gas}<7/2$, $x_0$ is the solution to
\begin{equation}
  t(x,x_0) =
  \frac{\Sigma_0\,\rc\,2^{(3-2\,p_\mathrm{gas})/4}\, \exp\left[-x_0/2\right]}
  {f_\mathrm{d}\,S\,\Sigd(x_0,0)\,\Vc \,  x_0^{(2\,p_\mathrm{gas}+1)/4}}\,
  \left(\Gamma\left[\frac{7-2\,p_\mathrm{gas}}{4},\frac{x}{2}\right]-
        \Gamma\left[\frac{7-2\,p_\mathrm{gas}}{4},\frac{x_0}{2}\right]\right),
  \label{eq:drift_characteristic}
\end{equation}
which needs to be solved numerically. We can find approximate solutions for $x_0$ by approximating 
the gamma function as before. In the case of $x_0,x\gg 1$, Eq.~\ref{eq:drift_characteristic} yields
\begin{equation}
x_0 = \frac{1}{2} \, \mathcal{W}\left[2 \left( f_\mathrm{d}\,S \,\epsilon \, \Omega_\mathrm{c}\,t
\right)^4 \frac{\exp(x)}{x} \right],
\end{equation}
where $\Omega_\mathrm{c}$ is the Keplerian frequency at \rc.
Eq.~\ref{eq:sigma_drift} also holds for a gas surface density profile
\begin{equation}
\Sigg(r) = \Sigma_\mathrm{c} \, \left(\frac{r}{\rc}\right)^{-p_\mathrm{gas}},
\label{eq:sig_drift_B}
\end{equation}
in which case $r_0$ is the solution of
\begin{equation}
t=\frac{2}{7-2\,p_\mathrm{gas}}\frac{\Sigma_\mathrm{c}\,r_0}{f_\mathrm{d}\,S\,\Sigd(r_0,0)\,\Vk(r_0)}\,\left(\frac{r_0}{\rc}\right)^{-p_\mathrm{gas}}\,\left[1-\left(\frac{r}{r_0}\right)^{(7-2\,p_\mathrm{gas})/4}\right].
\end{equation}

\section{E. Fragmentation-limited Solution}\label{app:frag_limited}
The Stokes number in the fragmentation-limited case (i.e., when the maximum impact velocity exceeds
the fragmentation velocity, $u_\mathrm{frag}\lesssim \sqrt{\alphat}\,\cs$) can be approximated as
\begin{equation}
\St \simeq \frac{1}{3\,\alphat}\,\frac{u_\mathrm{frag}^2}{\cs^2},
\end{equation}
which gives a drift velocity of 
\begin{equation}
u \simeq -B\sqrt{x}\left(x+\frac{q+5}{2}\right),
\end{equation}
with $B=f_\mathrm{f}\,u_\mathrm{frag}^2/3\alphat\,\Vc$ and $f_\mathrm{f}=0.37$
\citep[see][]{Birnstiel:2012p17135}. Using this velocity in Eq.~\ref{eq:t_integral}, we can
derive the solution for $x_0$ to be
\begin{equation}
x_0 =
\frac{q+5}{2}\,\left\{\tan\left[\frac{B}{2}\sqrt{\frac{q+5}{2}}\frac{t}{\rc}-\arctan\left(\sqrt{\frac{q+5}{2\,x}}\right)\right]\right\}^{-2},
\label{eq:frag_limit_x0}
\end{equation}
which, used in Eq.~\ref{eq:general_solution}, describes the time evolution of \Sigd in the 
fragmentation-limited case.

\section{F. Advection of a tracer in a viscously evolving accretion disk}\label{app:coadvection}
Quite similar to before (cf., Eq.~\ref{eq:sigg} and assuming $p_\mathrm{gas}=1$), the time
dependent gas surface density can be written as
\begin{equation}
\Sigg(r,t) = \Sigma_0(t) \, \left(\frac{r}{\rc(t)}\right)^{-1} \exp\left(-\frac{r}{\rc(t)}\right),
\label{eq:gas_timedependent}
\end{equation}
however the characteristic radius and the normalization are now time dependent,
\begin{equation}
\Sigma_0(t) = \frac{C}{3\pi\nu_1}\,\left(\frac{\rc(t)}{r_1}\right)^{-5/2},
\end{equation}
and
\begin{equation}
\rc(t) = r_1+ \frac{3\,t\,\nu_1}{r_1},
\end{equation}
as can be derived from \citet{Hartmann:1998p664} or \citet{LyndenBell:1974p1945}. $C$ is a constant
normalization factor proportional to the initial disk mass. The gas velocity then also becomes time-dependent,
\begin{equation}
u_\mathrm{gas}(r,t) = \frac{3 \nu_1}{r_1} \left(\frac{r}{\rc(t)}-\frac{1}{2}\right), 
\label{eq:u_gasdrag_timedep}
\end{equation}
and we assumed the viscosity to follow $\nu= \nu_1\cdot(r/r_1$). Small particles with $\St \ll 
1$ then move with the gas radial velocity, and the time evolution can be solved for using 
characteristics as before, yielding
\begin{equation}
\Sigd(r,t) = \Sigd(r_i,0)\cdot \left[
\left(\frac{\rc(t)}{r_1}\right)^2\,\left(1+\frac{1}{2}\,\frac{r_1}{r_i}\log\left(\frac{r_1}{\rc(t)}\right)\right)
\right]^{-1},
\label{eq:coadvection}
\end{equation}
where the initial position of a characteristic is defined to be
\begin{equation}
r_i(r,t) = \frac{r_1}{2}\cdot\left[\ln\left(\frac{\rc(t)}{r_1}\right)+2\,\frac{r}{\rc(t)}\right].
\label{eq:coadvection_ri}
\end{equation}
To semi-analytically calculate the evolution of \Sigd when both
radial drift (according to Appendix~\ref{app:drift_limited}) and viscous spreading are taken into
account, we iterated between these two analytical solutions:
\begin{enumerate}
  \item choose a time step $\Delta t$ within which the gas surface density does not change
  dramatically
  \item calculate how the dust is advected along with the gas according Eq.~\ref{eq:coadvection}
  during $\Delta t$
  \item use the result from (2) as an initial condition for Eq.~\ref{eq:sigma_drift} using the 
  same time step
  \item continue with step 1
  \end{enumerate}
  The results shown in Fig.~\ref{fig:comparison} were derived using nine logarithmically spaced
  time steps between $10^5$ and $10^6$ years.

\end{document}